\documentclass[twocolumn,aps,superscriptaddress,showpacs,preprintnumbers,floatfix,a4paper]{revtex4}

\usepackage{graphicx}
\usepackage{dcolumn}
\usepackage{bm}
\usepackage{color}
\usepackage{amssymb}
\usepackage{natbib}

\begin{document}

\title{Giant half-cycle attosecond pulses}
\author{H.-C. Wu}
\thanks{e-mail: hcwu@lanl.gov or huichunwu1@gmail.com}
\affiliation{Los Alamos National Laboratory, Los Alamos, New Mexico 87545, USA}
\author{J. Meyer-ter-Vehn}
\affiliation{Max-Planck-Institut f\"{u}r Quantenoptik, D-85748 Garching, Germany}
\date{\today }

\begin{abstract}
Half-cycle picosecond pulses have been produced from thin photo-conductors,
when applying an electric field across the surface and switching on
conduction by a short laser pulse. Then the transverse current in the wafer
plane emits half-cycle pulses in normal direction, and pulses of $500$ fs
duration and $10^{6}$ V/m peak electric field have been observed. Here we
show that single half-cycle pulses of $50$ as duration and up to $10^{13}$
V/m can be produced when irradiating a double foil target by intense
few-cycle laser pulses. Focused onto an ultra-thin foil, all electrons are
blown out, forming a uniform sheet of relativistic electrons. A second
layer, placed at some distance behind, reflects the drive beam, but lets
electrons pass straight. Under oblique incidence, beam reflection provides
the transverse current, which emits intense half-cycle pulses. Such a pulse may completely ionize even heavier
atoms. New types of attosecond pump-probe experiments will become possible.
\end{abstract}

\maketitle

\begin{figure}[t]
\includegraphics[width=0.4\textwidth]{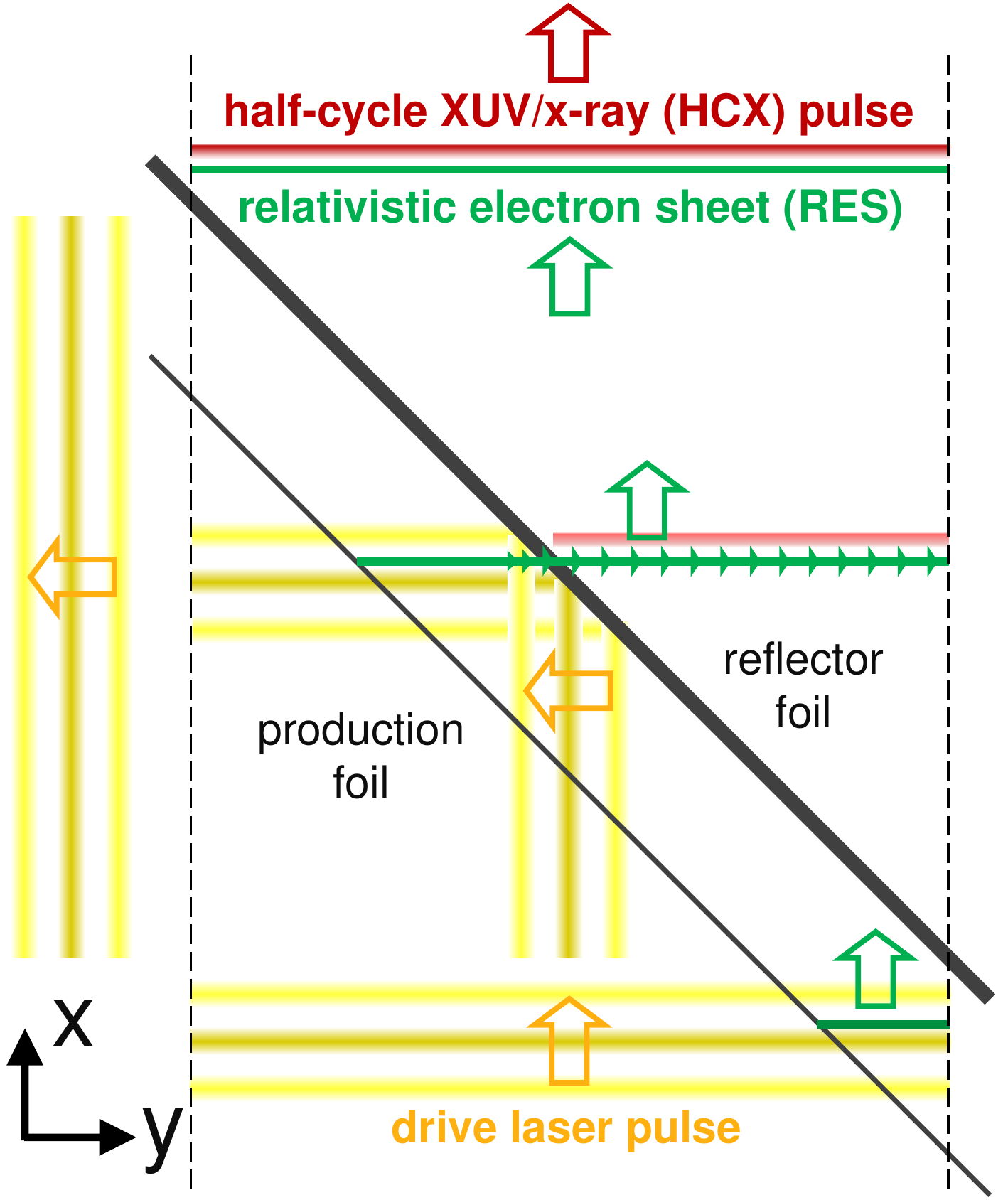} \label{fig1}
\caption{ \textbf{Scheme of target interaction and half-cycle emission.} A
three-cycle, $s$-polarized laser pulse (central intensity maxima shown in
yellow-brown) is obliquely incident on a double foil target. The drive laser
pulse is shown at different times with yellow arrows showing propagation
direction: (1) it starts passing the transparent electron production foil on
the right-hand side, ionizing and accelerating electrons which then move
(green arrows) as relativistic electron sheet (RES, green) parallel to the
wavefront ; (2) the drive pulse is reflected by the reflector foil, while
the electron sheet is moving straight on; due to conservation of canonical
momentum (see Methods section), electrons get a sudden transverse kick
(indicated by small green arrows); the corresponding unilateral current
radiates a short half-cycle XUV/x-ray (HCX) pulse (red) propagating just in
front of the electron layer; (3) the reflected laser pulse leaves to the
left, while the HCX followed by the RES exits vertically in the initial
direction of the drive pulse. }
\end{figure}

\begin{figure*}[t]
\includegraphics[width=1\textwidth]{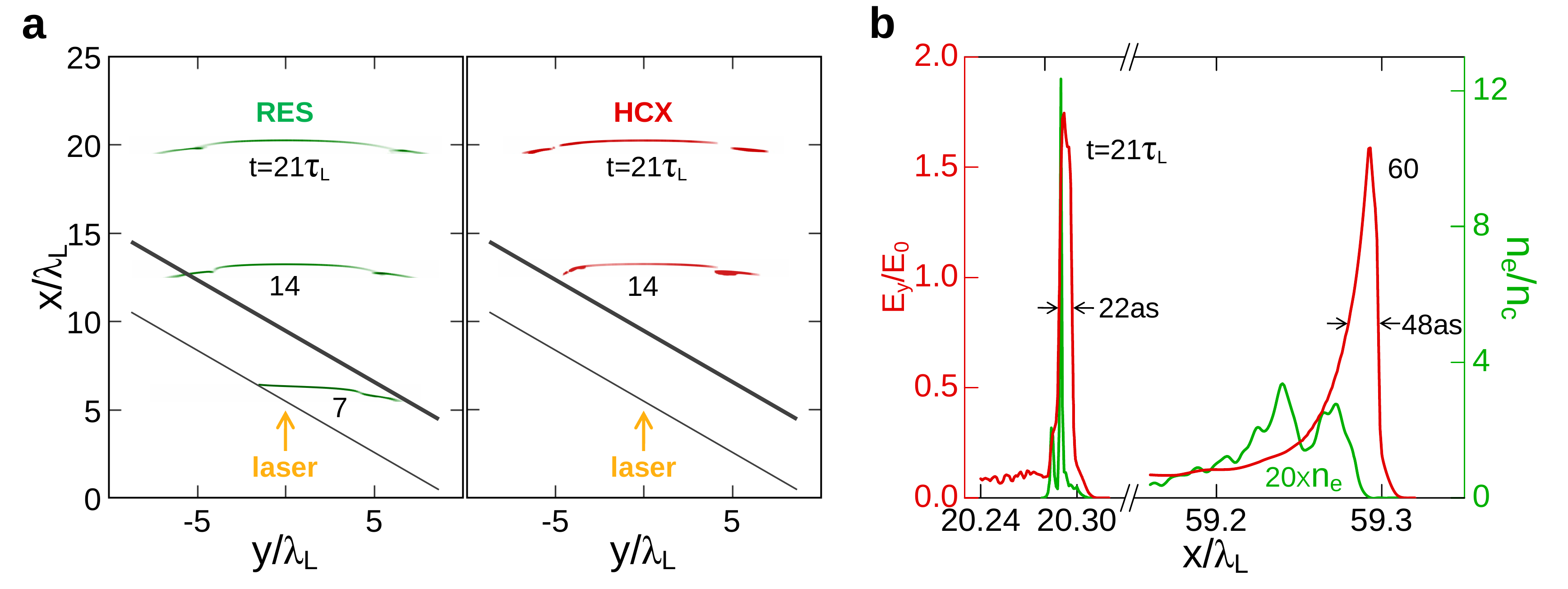} \label{fig2}
\caption{\textbf{Results of 2D PIC simulation.} \textbf{a}, Left panel:
snapshots of RES in $x,y$ plane plotted as electron density (green) at times
$7$, $14$, and $21\tau_L$; right panel: HCX plotted in terms of $%
E_y$ (red) at times $14$ and $21\tau_L$; the oblique lines
indicate production and reflector foil. \textbf{b}, Density $n_e$ (in units
of critical density $n_c=\protect\varepsilon_0m\protect\omega_L^2/e^2$) and
normalized electric field $a_y=E_y/E_0$ plotted versus $x$ along the central
line $y=0$ at times $21$ and $60\tau_L$. Simulation parameters are
given in the text and Methods section.}
\end{figure*}

\begin{figure*}[t]
\includegraphics[width=.8\textwidth]{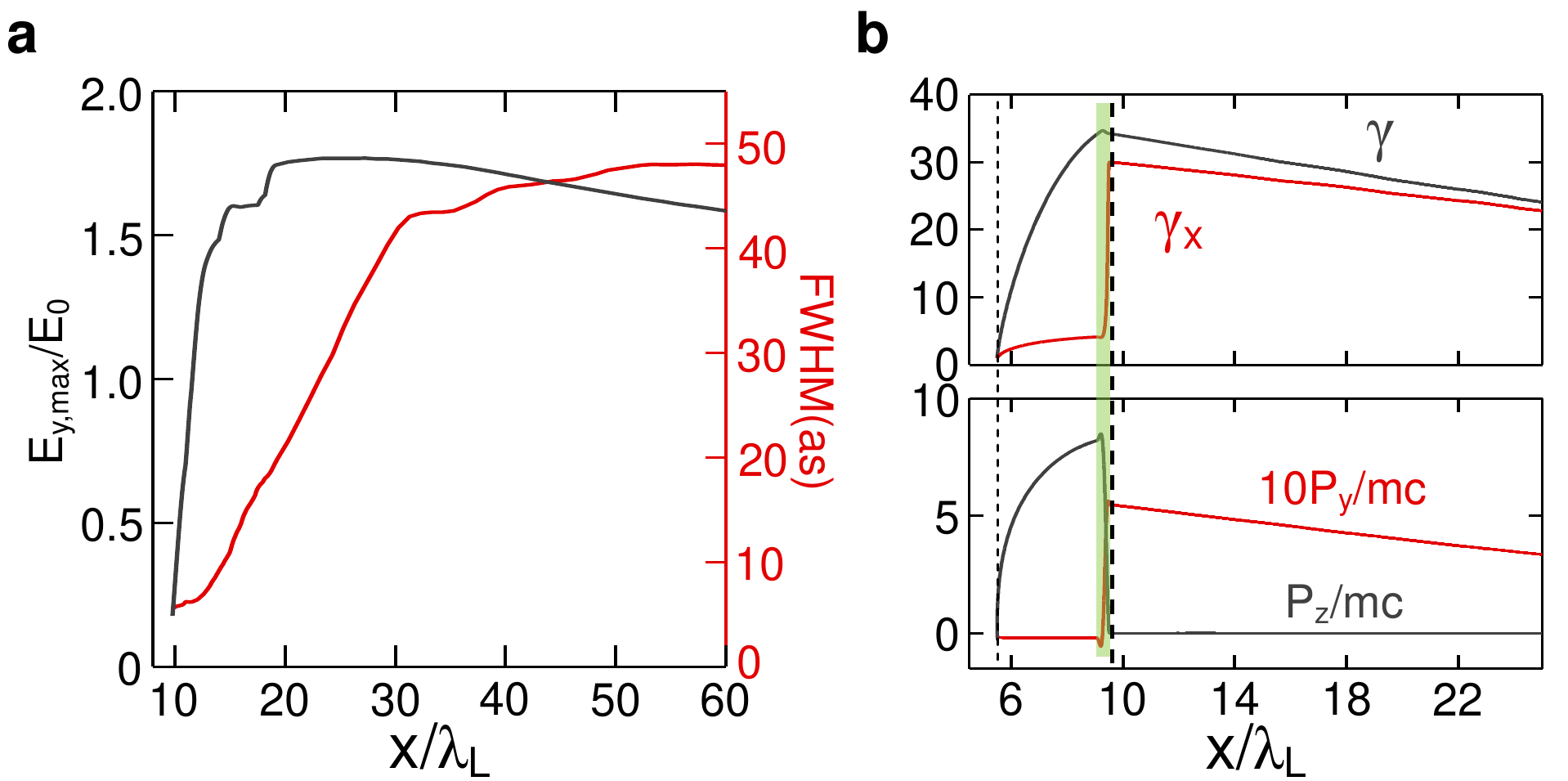} \label{fig3}
\caption{\textbf{Evolution of HCX and RES.} \textbf{a}, Evolution of peak electric field $E_{y,max}/E_{0}$
(black) and FWHM width (red) of the HCX pulse along the propagation
coordinate $x/\protect\lambda _{L}$. \textbf{b}, Averaged values of $\protect%
\gamma $ (black), $\protect\gamma _{x}$ (red) in upper panel, and $p_{y}/mc
$ (red, scaled by factor 10) and $p_{z}/mc$ (black) of the RES plotted
versus $x$ position. Averages are taken in $x$ direction along the $y=0$
line. Location of the production foil (thin dashed) is at $x/\protect\lambda %
_{L}=5.5$, and the reflector foil (thick dashed) is at $x/\protect\lambda %
_{L}=9.5$. In the shaded region, electron momenta change due to interaction
with the reflected drive laser.}
\end{figure*}

A major goal of attosecond science \cite{Krausz2009} is to follow electron
dynamics on the atomic time scale (24 as), be it in atoms or solids.
Different methods to produce single attosecond pulses to this end have been
described recently, and some of them have been tested experimentally \cite%
{Goulielmakis2008}. These methods rely on the generation of high harmonics
spectra which correspond to trains of as pulses in the time domain.
Generating these spectra with few-cycle drive pulses and applying high-pass
filters may be used to isolate single spikes. For pump-probe experiments,
sufficiently intense pulses are required, and surface harmonics generated at
solid-density plasma surfaces may provide high enough photon numbers \cite%
{Tsakiris2006}.

In the present letter, we follow a very different route, producing
half-cycle attosecond pulses from ultrathin relativistic electron sheets
(RESs). We note that similar half-cycle pulses, radiated from current sheets
in photo-conductors at rest, have actually been observed \cite{You1993} and
have been used in ionization experiments \cite{Jones1993}. Another proposal
is to produce disk-like, relativistic electron pulses from a synchrotron and
to kick them sideward by a magnetic field; picosecond 200 MW half-cycle
pulses are predicted this way \cite{Bratman2001}.

The present work is based on two recent developments: (1) The generation of
few-cycle high-contrast laser pulses at intensities exceeding $10^{19}$ W/cm$%
^{2}$ \cite{Herrmann2009}. Making use of OPCPA techniques \cite%
{Mikhailova2011} and plasma mirrors \cite{Dromey2006}, contrast ratios
beyond $10^{10}$ (subpicosecond time intervals as required in this case)
have been reached. High contrast is crucial to not destroy the ultra-thin
targets foils prematurely. (2) The second innovation concerns the
fabrication of few or even single carbon-atom layer graphenes \cite%
{Geim2007,Bai2010}. The graphene films are transparent to the laser light,
even at electron densities of $n_{e}\approx 10^{24}$ cm$^{-3}$ after full
ionization. An outstanding feature here is that laser pulses now available
can blow out all electrons from these foils. This happens when the laser
electric amplitude $E_{L}$ is larger than the charge separation field $%
E_{s}=en_{e}d/\varepsilon _{0}$, where $\varepsilon _{0}$ is the vacuum
dielectric constant, $d$ the thickness, and $en_{e}d$ the charge areal
density of the foil. It is the $v\times B_{L}$ part of the laser interaction
accelerates all electrons in laser direction. Relativistic energies up
to a maximum of $\gamma =1+a_{0}^{2}/2$ are reached over a few micrometer;
here $a_{0}=E_{L}/E_{0}$ is the normalized field strength, $E_{0}=mc\omega
_{L}/e$ the normalizing field, $\omega _{L}$ the laser circular frequency, $%
e $ and $m$ electron charge and mass, and $c$ the velocity of light. The
electrons then form a thin RES, separated from the ions and surfing on the
laser wave front \cite{MtV2009}.

These high-density sheets may serve as relativistic mirrors, compressing
femtosecond probe pulses to attosecond pulses by Doppler shifts of $4\gamma
^{2}$. For this to happen one has to divert the drive pulse from the RES by
an additional foil that is just thick enough ($\approx $ $50$ nm) to reflect
the light, but let the relativistic electrons pass almost unperturbed.
Behind the reflector, the electrons propagate in field-free ($a=0$) space,
where, due to conservation of canonical momentum, also their transverse
momentum ($p_{\perp }/mc=a$) vanishes. This means that the RES moves
precisely in normal direction, i.e. the direction of the incident drive
pulse. This makes it a perfect relativistic mirror for coherent Thomson
backscattering \cite{Wu2010}.

Here we consider the same configuration, but now for oblique incidence of
$s$-polarized laser light. A schematic drawing is given in Fig. 1 and a
corresponding two-dimensional (2D) particle-in-cell (PIC) simulation in Fig.
2. The amazing new feature is the appearance of a strong half-cycle
attosecond XUV/x-ray (HCX) electromagnetic pulse, preceding the RES after
emerging from the reflector layer. This is the central new result of this
paper. As we shall explain in the Methods section, the HCX pulse results
from conservation of canonical momentum under the condition of oblique
incidence. Due to interaction with the reflected laser pulse, each RES
electron gets a transverse kick ($p_{y0}=mc\tan \theta $) in the $y$
direction of the RES plane. The corresponding transverse current radiates
the unipolar half-cycle pulse.

In the simulated case, an $s$-polarized laser pulse is incident on target
under an angle of $\theta =30^{\circ }$. The laser pulse is given by $a(\tau
,y)=a_{0}\sin ^{2}(\pi \tau /T_{L})\exp (-y^{2}/R_{L}^{2})\cos (2\pi \tau
/\tau _{L})$ for $0<\tau <T_{L}$ with $\tau =t-x/c$, $a_{0}=20$, $\tau
_{L}=\lambda _{L}/c$, $T_{L}=3\tau _{L}$, and $R_{L}=5\lambda _{L}$. The
central intensity is $I_{L}=8.55\times 10^{20}$ W/cm$^{2}$ for wavelength $%
\lambda _{L}=800$ nm. The production layer has electron density $%
n_{e}/n_{c}=30$ and thickness $d/\lambda _{L}=0.001$, the reflector layer $%
n_{e}/n_{c}=500$ and thickness $d/\lambda _{L}=0.05$, where $%
n_{c}=1.74\times 10^{21}$ cm$^{-3}$ is the critical density for $\lambda
_{L}=800$ nm. The two layers are separated by $4\lambda _{L}$ in $x$
direction. The areal electron density of the production foil is $%
n_{e}d=4.17\times 10^{15}$ cm$^{-2}$, corresponding to nano-meshed graphene
\cite{Bai2010}. Other details on simulation setup are given in the Methods
section. Complete separation of electrons from ions creates an electrostatic
field of $E_{s}=0.76$ TV/m; it is about two orders of magnitude smaller than the laser
electric field of $E_{L}=80$ TV/m. In Fig. 2a, one sees how this laser pulse
drives the electrons out of the production foil (left panel) and how they
pass the reflector in $x$ direction and form of a thin sheet. RES density is
plotted for times $7$, $14$, and $21\tau _{L}$. The electron sheet is
accompanied by a strong unilateral $E_{y}$ pulse, plotted in the right panel
of Fig. 2a at times $14$ and $21\tau _{L}$.

Most important is Fig. 2b, giving line-outs of both electron density $n_{e}$
(green) and $E_{y}$ field (red) along the central line ($y=0$) at times $21$
and $60\tau _{L}$. It shows how the HCX pulse emerges and gradually
separates from the electron layer. The maximum field and the width
of the HCX pulse are also plotted in Fig. 3a, as they evolve along the
propagation coordinate. At $t=21\tau _{L}$, the HCX pulse has reached the
peak electric field of $E_{y}/E_{0}=1.75$, corresponding to $E_{y}=7$ TV/m
and a peak intensity of $I_{HCX}=1.3\times 10^{19}$ W/cm$^{2}$. This peak is
then slowly falling due to diffraction, while the HCX width is still
growing, saturating at $t=60\tau _{L}$. Plotted as electric field, the
FWHM width of the saturated pulse is 48 as, corresponding to about 24 as,
when plotted as intensity. This is the atomic time unit. At $t=60\tau _{L}$,
the relativistic electron sheet has strongly broadened due to Coulomb
expansion, and the peak density of initially $30n_{c}$ has fallen to $%
0.2n_{c}$. Now the HCX pulse has almost separated from the RES and is
propagating through vacuum. A remarkable feature is the very sharp front
edge, rising from zero to peak values within a few 10 attoseconds. In the
present simulation the pulse carries a total energy of $55$ $\mu $J, which
is about $10^{-4}$ of the incident laser energy. We have checked that
this pulse, about $7$ nm long, keeps its half-cycle character over almost a
millimeter, before it converts into a single-cycle pulse due to diffraction
loss of the long-wavelength components.

Let us now discuss the build-up of transverse current and HCX emission in
more detail. The reflector foil plays the key role in switching electron
momenta. This is documented in Fig. 3b. It displays $\gamma $, $\gamma _{x}$%
, $p_{y}/mc$, and $p_{z}/mc$ as function of RES position $x$ along the
central line $y=0$. The vertical dashed lines mark the positions of
production and reflector foil. It is seen that, during the initial
acceleration phase, RES electrons follow qualitatively relativistic single
electron motion, which satisfies $\gamma -1=p_{x}/mc=(p_{z}/mc)^{2}/2$ and $%
p_{y}/mc=0$ when driven by a planar laser pulse polarized in $z$ direction
\cite{MtV2001}. When passing the shaded region close to the reflector foil, $%
p_{z}$ falls to zero, as the laser amplitude $a_{z}$ does, and $p_{y}$ pops
up, approaching the predicted value of $p_{y0}/mc=\tan \theta =0.577$. This
behavior is derived in the Methods section. Inside the dense reflector
plasma, the transverse current due to $p_{y}$ is screened, but as soon as it
emerges from the rear side, it radiates the HCX pulse, while $p_{y}$
decreases due to radiation damping. At this point, the present case differs
from the photo-conductor pulses, which are mainly emitted during the
build-up phase of the transverse current and have opposite polarity. In Fig.
3b, we have also plotted $\gamma _{x}=(1-\beta _{x}^{2})^{-1/2}$, which is
related to the full $\gamma $ by $\gamma =\gamma
_{x}[1+(p_{y}/mc)^{2}+(p_{z}/mc)^{2}]^{1/2}$. One might expect that $\gamma
_{x}$ stays constant behind the reflector, because there is no further
acceleration by the drive laser. However, there is deceleration due to the
charge separation field $E_{s}$, which still acts behind the reflector. In
particular, tail electrons facing the reflector are decelerated, while front
electrons cruise at constant speed. This causes expansion, and a $\gamma
_{x} $ distribution develops over the sheet. What is actually plotted in
Fig. 3b are averages over these distributions.

The electron sheet, moving with $\gamma _{x}$ in $x$ direction and having
superimposed transverse momentum $p_{y}$, emits electromagnetic radiation
proportional to the time derivative of the transverse current density in the
sheet. In the Methods section, the basic equation is solved for the
idealized case of a uniform zero-thickness RES. Of course, such treatment
does not account for finite-size RES dynamics, but provides useful estimates
and scalings. The HCX peak electric field is obtained as $E_{max}=\gamma
_{x0}E_{s}\sin \theta =11$ TV/m for the reference case with the separation
field $E_{s}=0.76$ TV/m and $\gamma _{x0}=30$ (see Fig. 3b). It should be
compared with the simulated value $E_{max}=7$ TV/m at $t=21\tau _{L}$. In
zero-thickness approximation, the HCX has an infinitely sharp front and then
decays exponentially with time constant $T\approx mc/(\gamma _{x0}eE_{s})=75$
as, corresponding to a FWHM width of $52$ as, which is close to the
48 as, found in Fig. 2b.

Finer details of the HCX front edge and also the HCX growth along the
propagation axis, observed in Fig. 3a, are related to the finite size of the
RES, when emerging from the reflector. In the simulated density profile at
this time, the RES shows two peaks at a distance of $\Delta x_{1}\approx
0.008\lambda _{L}$ and some precursor foot extending over $\Delta
x_{2}\approx 0.03\lambda _{L}$. The precursor maps directly into the HCX
front edge rising over almost the same distance $\Delta x_{2}$. Also the
propagation distance for saturating the peak HCX field can be estimated as $%
L_{sat}=\Delta x_{1}/(1-\beta _{x})\approx 2\gamma _{x}^{2}\Delta
x_{1}\approx 14\lambda _{L}$, in fair agreement with Fig. 3a. Here $L_{sat}$
is the distance over which the light signal emitted from the second density
peak has to travel to catch up with the first peak. These two major HCX
contributions then add up coherently and form the peak of the pulse. Due to
retardation the front contribution is already somewhat damped, and this may
explain why the peak field of $7$ TV/m is lower than the model result of $11$
TV/m. It is important to notice that the HCX contributions emitted from
different RES layers add up coherently. The coherence is the reason why a
major part of the energy deposited in the transverse RES current is actually
radiated, even when the RES has substantially expanded.

In conclusion, we have found an efficient new option to generate single
attosecond pulses. It involves a double foil target to produce and purify a
RES. Under oblique incidence, it radiates a half-cycle pulse of a few 10 as
duration with peak electric field up to $~10^{13}$ V/m and pulse energies up
to the $0.1$ mJ level. The laser-to-HCX conversion efficiency amounts to a
few $10^{-4}$. Most important for experiments is the extremely sharp front
edge with rise times of a few $10$ as. HCX properties and scaling relations
have been derived in simple analytical terms. Compared to Thomson scattering
as a way to generate single attosecond pulses \cite{Wu2010}, the present
method is simpler to implement, because only a single laser beam is
required. It is hoped that the present paper will stimulate experiments.


\begin{figure}[t]
\includegraphics[width=0.45\textwidth]{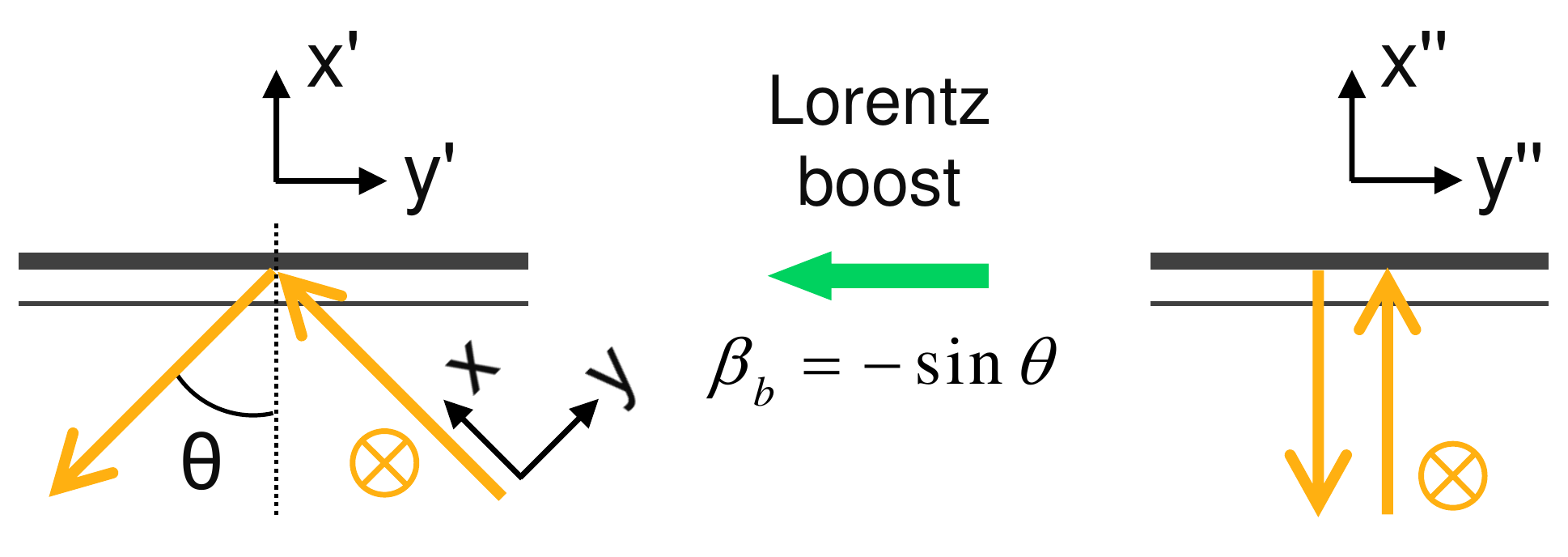} \label{fig4}
\caption{\textbf{Definition of coordinate transformations.}}
\end{figure}

\bigskip \textbf{\large Methods}

\textbf{Transverse momentum.} For oblique incidence, electron dynamics close
to the reflector are quite complex. A simple analytic approach is possible
when transforming to a frame in which the laser beam is vertically incident
as depicted in Fig. 4 \cite{Bourdier1983}. First rotating the $x,y$ beam
frame to the $x^{\prime },y^{\prime }$ layer frame and then applying a
Lorentz-boost in $y^{\prime }$ direction with $\beta _{b}=-\sin \theta $, $%
\gamma _{b}=\sec \theta $, electron energy-momentum $(\gamma ,\hat{p}_{x},%
\hat{p}_{y},\hat{p}_{z})$ transforms to $(\gamma ^{\prime \prime },\hat{p}%
_{x}^{\prime \prime },\hat{p}_{y}^{\prime \prime },\hat{p}_{z}^{\prime
\prime })$ with $\gamma ^{\prime \prime }=(1+\hat{p}_{x}^{\prime \prime
}{}^{2}+\hat{p}_{y}^{\prime \prime }{}^{2}+\hat{p}_{z}^{\prime \prime
}{}^{2})^{1/2}$; here energy is in units $mc^{2}$ and momenta in $mc$. Due
to conservation of canonical momentum, the $z$-component of momentum
(vertical to the plane of Fig. 4) is invariant and equals the vector
potential $a_{z}$ for s-polarized light, $\hat{p}_{z}=\hat{p}_{z}^{\prime }=%
\hat{p}_{z}^{\prime \prime }=a_{z}$. In the boosted frame, also the $y$%
-component $\hat{p}_{y}^{\prime \prime }=\beta _{b}\gamma _{b}=\tan \theta $
is conserved. Transforming back to the lab frame, we find
\begin{equation}
\hat{p}_{y}=\hat{p}_{y}^{\prime \prime }+(\hat{p}_{x}^{\prime \prime
}-\gamma ^{\prime \prime })\sin \theta
\end{equation}%
with $\hat{p}_{x}^{\prime \prime }-\gamma ^{\prime \prime }=\hat{p}%
_{x}^{\prime \prime }-(1+\tan ^{2}\theta +\hat{p}_{x}^{\prime \prime
}{}^{2}+a_{z}^{2})^{1/2}$. Here we notice that, behind the reflector where $%
a_{z}=0$, we have $\hat{p}_{x}^{\prime \prime
}-\gamma ^{\prime \prime }\rightarrow 0$ for relativistic electrons, provided that $\hat{p}%
_{x}^{\prime \prime }\cos \theta \gg 1$. In this case, one has $\hat{p}%
_{y}\approx \hat{p}_{y}^{\prime \prime }=\tan \theta $, stating that
electrons emerge from the reflector with non-zero momentum in the direction
of the RES plane. Momentum transfer into $y$-direction occurs within a
narrow region close to the reflector (see Fig. 3b), where $a_{z}\rightarrow
0 $.

\bigskip \textbf{Scaling of HCX emission.} HCX generation from a RES can be
described by the one-dimensional wave equation
\begin{equation}
(\partial ^{2}/\partial x^{2}-c^{-2}\partial ^{2}/\partial
t^{2})E_{y}=\varepsilon _{0}^{-1}c^{-2}\partial J_{y}/\partial t,
\end{equation}%
where $J_{y}=-ec\beta _{y}n_{e}$ is the radiating current. The solution can
be written as $E_{y}(x,t)=-(2\varepsilon _{0}c)^{-1}\int \int dx^{\prime
}dt^{\prime }H(t-t^{\prime }-|x-x^{\prime }|/c)\partial J_{y}(x^{\prime
},t^{\prime })/\partial t^{\prime }$, where $H(t)$ is the step function. For
analytical treatment, the difficulty is that $J_{y}(x,t)$ is a complicated
function of space and time in general. Here we restrict ourselves to a RES
of zero-thickness, approximating the density profile as $n(x)=n_{e}\delta
(x/d)$ with finite areal density $\int n(x)dx=n_{e}d$, and first consider
the rest frame (index R), choosing uniform velocity $\beta _{xR}=0$ and $%
\beta _{yR0}=\sin \theta $ initially. The solution consists of two
half-cycle electromagnetic pulses emitted symmetrically in $\pm x$
direction, just as observed in the photon-conductor experiments \cite%
{You1993}. Expressing $J_{y}$ in terms of transverse momentum $p_{y}$, we
find for the wave traveling in $+x$ direction
\begin{equation}
E_{yR}(\tau _{R})=(E_{s}/2)p_{y}(\tau _{R})/(mc\gamma _{y}),
\end{equation}%
where $\tau _{R}=t-x/c$ and $E_{s}=en_{e}d/\varepsilon _{0}$. The
relativistic factor $\gamma _{y}=\sqrt{1+(p_{y}/mc)^{2}}$ can be estimated
as $\gamma _{y}\approx 1$. At $x=0$ (i.e. $\tau _{R}=t$), where the layer is
located, $E_{yR}$ damps electron momentum according to $%
dp_{y}/dt=-eE_{yR}(t) $. One finds that both $p_{y}(t)$ and the emitted
pulse $E_{yR}(\tau _{R})$ decay exponentially on time scale $%
T_{R}=2mc/(eE_{s})$. These results apply to the rest frame of the layer.
Performing a Lorentz-transformation to the lab frame, in which the layer is
moving with velocity $c\beta _{x}$ and $\gamma _{x}=(1-\beta
_{x}^{2})^{-1/2} $ in normal direction, we find the coordinate of the
forward HCX pulse $\tau =\gamma _{x}(1-\beta _{x})\tau _{R}\approx \tau
_{R}/(2\gamma _{x})$ and the electric field $E_{y}=\gamma _{x}(1+\beta
_{x})E_{yR}\approx 2\gamma _{x}E_{yR}$. The maximum HCX field scales like $%
E_{max}\propto \gamma _{x}E_{s}\sin \theta $ and the pulse width like $%
T\propto (\gamma _{x}E_{s})^{-1}$.

\bigskip \textbf{PIC simulation.} We have carried out all simulations using
the JPIC code \cite{WuArxiv}, which employs a field solver free of numerical
dispersion in $x$ direction \cite{Sentoku2008}. Since an isolated HCX pulse
contains extremely broad frequency components, for accurately simulating HCX
generation and propagation, it is crucial to use such kind of
dispersion-free field solver. The simulation box has a size $25\lambda
_{L}\times 20\lambda _{L}$ in the $xy$ plane. Moving window technique is
used to extend the simulation distance. There are 1000 cells and 400 cells
per $\lambda _{L} $ in $x$ and $y$ directions, respectively. Along the
central line $y=0$, the undersides of both foils are at $x=5.5\lambda _{L}$
and $9.5\lambda _{L}$, respectively. We use $3\times 10^{7}$ and $1\times
10^{7}$ macro-particles in the production and reflector foils, respectively.
Ions in both layers are immobile. An initial electron temperature of $10 $
eV is taken for the production sheet to mimic ionization, while the
reflector foil is initialized as cold plasma with zero temperature. The
drive laser begins to enter the simulation box from the lower boundary ($x=0$%
) at $t=0$.


\bigskip
\textbf{\large Acknowledgments}

This work is supported by the LDRD Program 20110341ER at the Los Alamos
National Laboratory. J. Meyer-ter-Vehn was supported by the Munich Center
for Advanced Photonics and by the Association EURATOM-Max-Planck-Institute
for Plasma Physics. H.-C. Wu is grateful for supports from Dr. J. Fernandez and
Dr. B.M. Hegelich.

\bigskip
\textbf{\large Author contributions}

H.-C.W. discovered the new effect described in this paper, he carried out all
simulations and developed the basic theory. J.M.-t.-V. wrote the paper and
clarified some details of the physics. Both authors take full responsibility for the results presented.

\bigskip
\textbf{\large Additional information}

The authors declare no competing financial interests. 
Correspondence and requests for materials should be addressed to H.-C.W.

\end{document}